\documentclass[twocolumn,showpacs,preprintnumbers,amsmath,amssymb]{revtex4}
\usepackage{graphicx}

\begin{document}
\title{Thermoelectric properties of a weakly coupled quantum dot: Enhanced thermoelectric efficiency}
\author{M. Tsaousidou}
\email{rtsaous@upatras.gr}
\affiliation{Materials Science
Department, University of Patras, Patras 26504, Greece}
\author{G.P. Triberis}
\affiliation{Physics Department, Solid State Section, University
of Athens, Panepistimiopolis, 15784, Zografos, Athens, Greece}

\begin{abstract}
We study the thermoelectric coefficients of a multi-level quantum
dot (QD) weakly coupled to two electron reservoirs in the Coulomb
blockade regime. Detailed calculations and analytical expressions
of the power factor and the figure of merit are presented. We
restrict our interest to the limit where the energy separation
between successive energy levels is much larger than the thermal
energy (i.e., the quantum limit) and we report a giant enhancement
of the figure of merit due to the violation of the Wiedemann-Franz
law when phonons are frozen. We point out the similarity of the
electronic and the phonon contribution to the thermal conductance
for zero dimensional electrons and phonons. Both contributions
show an activated behavior. Our findings suggest that the control
of the electron and phonon confinement effects can lead to
nanostructures with improved thermoelectric properties.
\end{abstract}

 \maketitle

\section{Introduction} The thermoelectric properties of
nanomaterials have attracted a great deal of interest in the last
few years due to the potential applications in power generation
and refrigeration \cite{Dresselhaus}. There is increasing evidence
that the thermoelectric performance of nanostructured materials
can be significantly improved. The efficiency of a device to
convert heat into electricity and vise-versa is measured by the
dimensionless figure of merit $ZT$ given by
\begin{equation}
ZT=\frac{S^{2}G T}{\kappa},
\end{equation}
where $S$ is the thermopower, $G$ is the electrical conductance,
$T$ is the absolute temperature and $\kappa$ is the thermal
conductance. There are two contributions to $\kappa$: the
electronic, $\kappa_{e}$, and the phonon, $\kappa_{ph}$.

Good thermoelectric materials are considered to be those with
$ZT>3$ at room temperature \cite{Majumdar}. The best commercial
thermoelectric material is bulk Bi$_{2}$Te$_{3}$ and its alloys
with Sb and Se with $ZT$ close to 1. Although there is no
thermodynamic requirement that imposes an upper limit on $ZT$,
until 2001 no progress had been made in increasing $ZT$ above 1.
The difficulty is due to the interrelations between the
coefficients $S$, $G$ and $\kappa$ that set limitations in varying
them independently. A characteristic example is the
Wiedemann-Franz law according to which the ratio $\kappa_{e}/G T$
remains constant.

Since 2000 there has been a number of experiments in
nanostructures where the barrier of $ZT=1$ has been overcome at
high temperatures. Venkatasubramanian {\em et al.} \cite{Venka}
measured $ZT\approx 2.4$ in $p$-type
Bi$_{2}$Te$_{3}$/Sb$_{2}$Te$_{3}$ superlattices at $T=300$~K. The
reason for the enhanced figure of merit is the phonon blocking due
to acoustic mismatch between Bi$_{2}$Te$_{3}$ and Sb$_{2}$Te$_{3}$
that causes a significant reduction to $\kappa_{ph}$. Harman {\em
et al.} \cite{Harman} reported values of $ZT$ between 1.3 to 1.6
in PbSeTe/PbTe quantum dot superlattices. More recently Kanatzidis
and co-workers found that bulk AgPb$_{18}$SbTe$_{20}$ with
internal nanostructures has $ZT\approx 2$ at $T=800$~K \cite{Hsu}.
In nanocrystalline BiSbTe bulk alloys $ZT$ reached the value 1.4
at $T=373$~K \cite{Poudel}. The enhanced $ZT$ was attributed to
the significant reduction of $\kappa_{ph}$ due to the strong
phonon scattering by the interfaces between the nanostructures. A
100-fold improvement of $ZT$ compared to the bulk value has been
reported recently in Si nanowires \cite{Hochbaum,Boukai}. In
\cite{Hochbaum} the increased $ZT$ is due to particularly small
values of $\kappa_{ph}$ that reach the amorphous limit, while in
\cite{Boukai} the phonon-drag effect \cite{Fletcher,Tsaousidou}
was identified as the reason for the increased thermopower at
$T=200$~K.

Early theoretical work pointed out the possibility of enhancing
$ZT$ in structures of reduced dimensionality due to the increased
density of states near the Fermi level \cite{Hicks1}. Mahan and
Sofo~\cite{Mahan} predicted a maximization of the thermoelectric
efficiency in materials with a delta-like density of states. This
suggests that quantum dots and molecular junctions are promising
candidates for good thermoelectric materials. Humphrey and Linke
\cite{Humphrey} provided a thermodynamic explanation why the
optimum density of states for enhanced ZT is a delta function.
More recent theoretical studies
\cite{Tsaousidou1,Tsaousidou2,Murphy,Lambert,Swirkowicz} point out
the possibility of significantly improving the thermoelectric
efficiency in small quantum dots and molecules.

The present work presents a comprehensive analysis of the
thermoelectric properties of a weakly coupled multi-level quantum
dot (QD) in the sequential tunneling transport regime and suggests
the possibility of a huge increase of $ZT$ in the Coulomb blockade
regime due to the violation of the Wiedemann-Franz law. We focus
on the quantum limit where the energy separation between
successive electronic levels is large compared to the thermal
energy. Preliminary results have been presented in a conference
form \cite{Tsaousidou2}. Deviations from the Wiedemann-Franz law
in zero dimensional structures were predicted also in
[\onlinecite{Zianni2,Kubala}] without, however, relating these to
the possibility of an enhancement of ZT. Recently, an interesting
prediction for a significant increase of $ZT$ in a single-level
molecule was made by Murphy {\em et al.}~\cite{Murphy}. In
Ref.~[\onlinecite{Murphy}] it was found that the energy parameter
that controls $ZT$ is the electron-electron repulsion. Here we
show that for a multi-level system in the Coulomb blockade regime
$ZT$ is controlled by the level spacing and, consequently, the
spatial confinement.

The Coulomb blockade oscillations of the conductance and
thermopower in the sequential tunneling regime have been studied
in the seminal papers by Beenakker \cite{Been1} and by Beenakker
and Staring\cite{Been2}. Recently, the theory in
[\onlinecite{Been1,Been2}], that is based on a master equation
approach within the constant interaction model, has been extended
to calculate the electronic contribution to thermal conductance
\cite{Tsaousidou1,Tsaousidou2,Zianni2}. Here we present, for the
first time, a detailed derivation for an analytical expression for
$\kappa_{e}$ which serves as a solid framework for the
understanding of the behavior of this coefficient in the quantum
limit. Within this framework we discuss the validity of previous
results \cite{Zianni2,Zianni1}. The cases of a QD with equidistant
and non-equidistant energy spectrum are examined. We also present
detailed calculations and analytical derivations for the power
factor $S^{2}G$ and the figure of merit. We predict an
optimization of the thermoelectric efficiency at specific values
of the Fermi level. An important outcome of this work is the
similarity in the temperature dependence of the electronic and
phonon contribution to $\kappa$ in structures with zero
dimensional (0D) electrons and phonons. Our findings suggest the
possibility of a dramatic increase in $ZT$ in {\em both} electron-
and phonon- blocking devices.

\section{Theory}
We consider a quantum dot that consists of non-degenerate discrete
energy levels $\epsilon_{n}$ ($n=1,2,3,...$) weakly coupled to two
electron reservoirs. In equilibrium both reservoirs have the same
temperature $T$ and the same chemical potential $E_{F}$. The
energy distribution in the reservoirs is the Fermi-Dirac function
$f(E-E_{F})=1/\{\exp[\beta(E-E_{F})]+1\}$, where
$\beta=(k_{B}T)^{-1}$. We examine the Coulomb blockade regime
where the charging energy $E_{C}=e^{2}/2C$ ($C$ is the capacitance
of the dot to the surroundings) is larger than the energy spacing
$\Delta \epsilon$ between successive energy levels. The tunnel
rate from level $n$ to the left, L, (right, R,) reservoir is
denoted as $\Gamma_{n}^{L}$ ($\Gamma_{n}^{R}$). We assume that
$k_{B}T$ and $\Delta\epsilon$ are much larger than
$h(\Gamma_{n}^{L}+\Gamma_{n}^{R})$ (where $h$ is Planck's
constant) and we neglect effects due the virtual tunneling
(cotunneling) of electrons through the dot \cite{Been1,Been2}.

The results presented here are based on the calculation of the
heat current $Q$ through the dot when small temperature and
voltage differences $\Delta T$ and $\Delta V$, respectively, are
applied between the two reservoirs. The linear response model
proposed in the pioneer work by Beenakker \cite{Been1} and
Beenakker and Staring \cite{Been2} for the calculation of the
charge current $J$ has been generalized to include $Q$
\cite{Tsaousidou1,Tsaousidou2,Zianni2}. Then $Q$ is given by
\begin{equation}
\label{heat}
Q=-\frac{\Gamma^{L}\Gamma^{R}}{\Gamma^{L}+\Gamma^{R}}\left
(s_{1}\frac{e\Delta V}{k_{B}T}+s_{2}\frac{\Delta
T}{k_{B}T^{2}}\right).
\end{equation}
In the above equation the energy dependence of the tunnel rates
has been ignored. $e$ is the magnitude of the electron charge.
Moreover, for the quantities $s_{m}$ with $m=0,1,2$ we use the
following notation
\begin{equation}
\label{sm}
s_{m}=\sum_{n=1}^{\infty}\sum_{N=1}^{\infty}P_{eq}(N)P_{eq}(\epsilon_{n}|N)
[1-f(E-E_{F})](E -E_{F})^{m}
\end{equation}
where, the sum indexes $n$ and $N$ refer to the energy level and
the number of electrons in the dot, respectively, $\epsilon_{n}$
is the energy of the $n$ state and $E=\epsilon_{n}+U(N)-U(N-1)$
with $U(N)$ being the electrostatic energy for a dot with $N$
electrons. Moreover, $P_{eq}(N)$ is the probability that the dot
contains $N$ electrons in equilibrium and $P_{eq}(\epsilon_{n}|N)$
is the probability that in equilibrium the $n$ level is occupied
given that the dot has $N$ electrons. The expressions for
$P_{eq}(N)$ and $P_{eq}(\epsilon_{n}|N)$ are given in
[\onlinecite{Been2}].

Utilizing the standard transport equation \cite{Tsaousidou}
\begin{equation}
Q=M\Delta V+K\Delta T
\end{equation}
the transport coefficients $M$ and $K$ can be readily obtained
\cite{Tsaousidou2}, namely,
\begin{equation}
\label{M} M=-\frac{e}{k_{B}T}\Gamma^{tot}s_{1}
\end{equation}
and
\begin{equation}
\label{K} K=-\frac{1}{k_{B}T^{2}}\Gamma^{tot}s_{2}.
\end{equation}
In the above equations
$\Gamma^{tot}=\Gamma^{L}\Gamma^{R}/(\Gamma^{L}+\Gamma^{R})$.

Onsager's symmetry relation relates the thermoelectric coefficient
$M$ to the thermopower $S$ by M=SGT \cite{Tsaousidou}. Then $S$ is
written as
\begin{equation}
\label{S} S=\frac{M}{GT}=-\frac{1}{eT}\frac{s_{1}}{s_{0}},
\end{equation}
where $G$ is given by \cite{Been1}
\begin{equation}
\label{G}
G=G_{0}s_{0},
\end{equation}
with $G_{0}=(e^{2}/k_{B}T)\Gamma^{tot}$. The expression (\ref{S})
we derive for $S$ is identical to the expression (3.13) obtained
by Beenakker and Staring \cite{Been2}, where the authors
calculated the electron current through the dot instead of the
heat current. The approach we use here for $S$ is the so-called
$\Pi$-approach \cite{Tsaousidou}.

Finally, the electronic contribution to the thermal conductance
can be calculated by using the standard relation~\cite{Tsaousidou}
\begin{equation}
\label{kappa} \kappa_{e}=-K-S^{2}GT.
\end{equation}

In Fig.1a we show the calculations of the thermoelectric
coefficient $M$ as a function of the Fermi level for a three-level
QD when $\beta\Delta\epsilon=10$ and $\Delta\epsilon=0.5 E_{C}$.
$M$ exhibits Coulomb blockade oscillations with periodicity
$\Delta E_{F}=\Delta \epsilon+(e^{2}/C)$ as does the electrical
conductance \cite{Been1}. The effect of temperature on the shape
of $M$ is shown in Fig.1b. The Coulomb blockade oscillations of
the coefficient $K$ are shown in Fig.2a for the same parameters.
As we can see $-K$ shows a double peak structure which is
reproducible in energy intervals $\Delta \epsilon+(e^{2}/C)$. The
theoretical values for $K$ have been divided by $L_{0}TG^{max}$
where $L_{0}=(k_{B}/e)^{2}\pi^{2}/3$ and $G^{max}=G_{0}/4$ is the
maximum value of the electrical conductance. The effect of
temperature on the peaks of $-K$ is shown in Fig.2b and it will be
explained by a simple analytical expression derived later.

In Fig.2 we have also shown the calculation of $S^{2}GT$. The
curves for $-K$ and $S^{2}GT$ are {\em indistinguishable} so that
$\kappa_{e}$ is near zero. It is found that the magnitude of the
maximum values of $\kappa_{e}$ collapses rapidly to zero as
$(\beta\Delta\epsilon)^{2}\exp(-\beta\Delta\epsilon)$ violating
the Wiedemann-Franz law. Usually, in degenerate 2D and 1D systems
$S^{2}GT$ is a small quantity compared to $-K$ and the thermal
conductance follows closely $-K$. The standard Sommerfeld
expansion leads to the Wiedemann-Franz law according to which
$\kappa_{e}=-K=L_{0}GT$. The above imposes a significant
limitation on ZT when $S^{2}<L_{0}$ even when the phonon
contribution to the thermal conductance is ignored.

\begin{figure}
\begin{centering}
\includegraphics[angle=0,height=9.5cm]{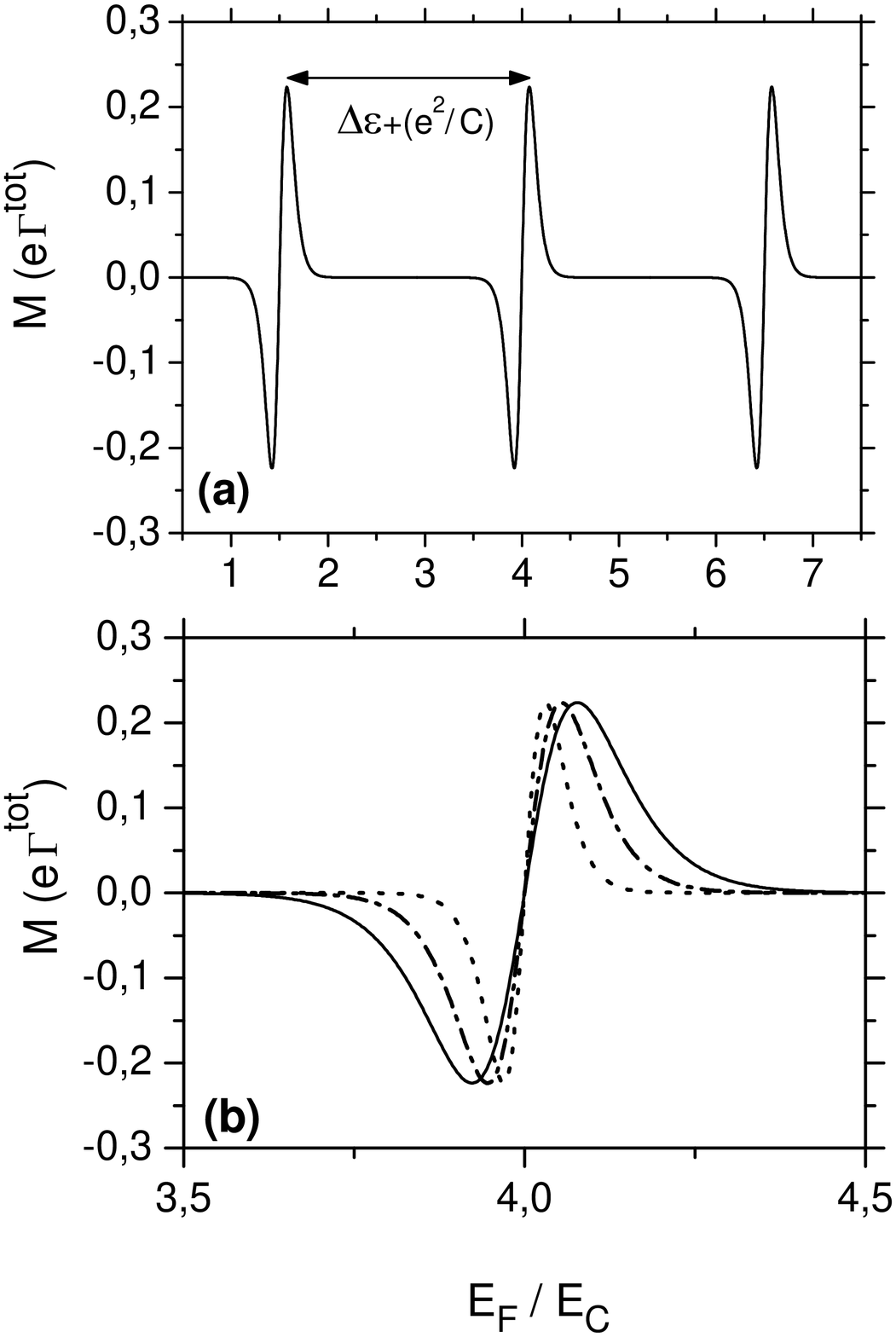}
\par\end{centering}

\caption{The Coulomb blockade oscillations of $M$ and the effect
of temperature on the shape of $M$ for a three-level dot. The
solid line in (a) is for $k_{B}T=0.05 E_{C}$. In (b) The solid,
dashed-dotted and dotted lines correspond to $k_{B}T=0.05,0.035$
and $0.02 E_{C}$, respectively. The level spacing is $\Delta
\epsilon=0.5 E_{C}$.}
\end{figure}

\begin{figure}
\begin{centering}
\includegraphics[angle=0,height=9.5cm]{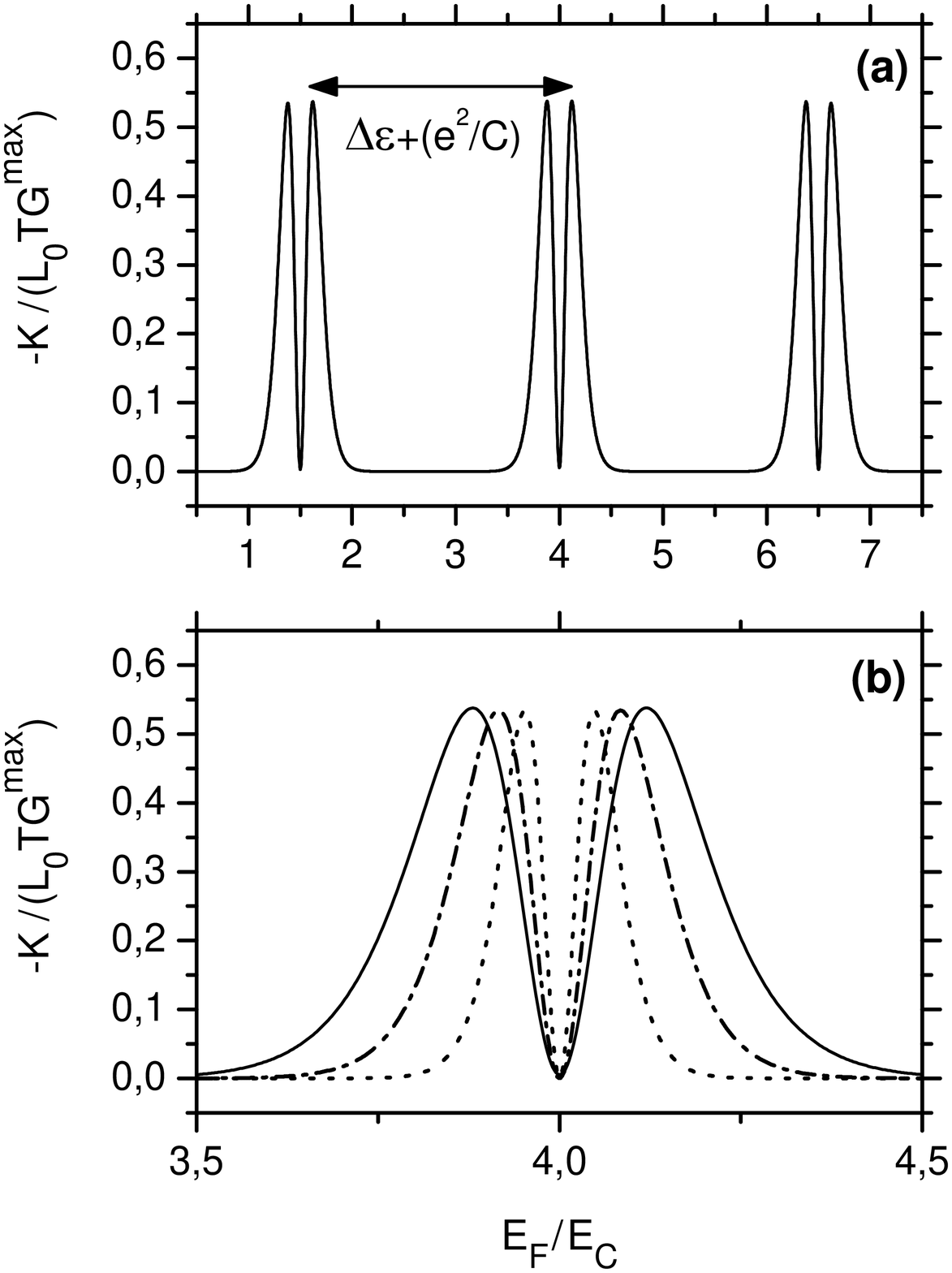}
\par\end{centering}

\caption{The coefficient $K$ as a function of $E_{F}$ for a
three-level dot. The calculated values of $K$ are divided by
$L_{0}TG^{max}$ where $G^{max}=G_{0}/4$ is the maximum value of
the conductance. In (a) the Coulomb blockade oscillations of $K$
are shown for $k_{B}T=0.05 E_{C}$. The effect of temperature on
the shape of $K$ is shown in (b). The solid, dashed-dotted and
dotted lines correspond to $k_{B}T=0.05,0.035$ and $0.02 E_{C}$,
respectively. The level spacing is $\Delta \epsilon=0.5 E_{C}$. In
the same figure we plot $S^{2}GT$. The curves for $-K$ and
$S^{2}GT$ are indistinguishable.}
\end{figure}

\begin{figure}
\begin{centering}
\includegraphics[angle=-90,width=7.3cm]{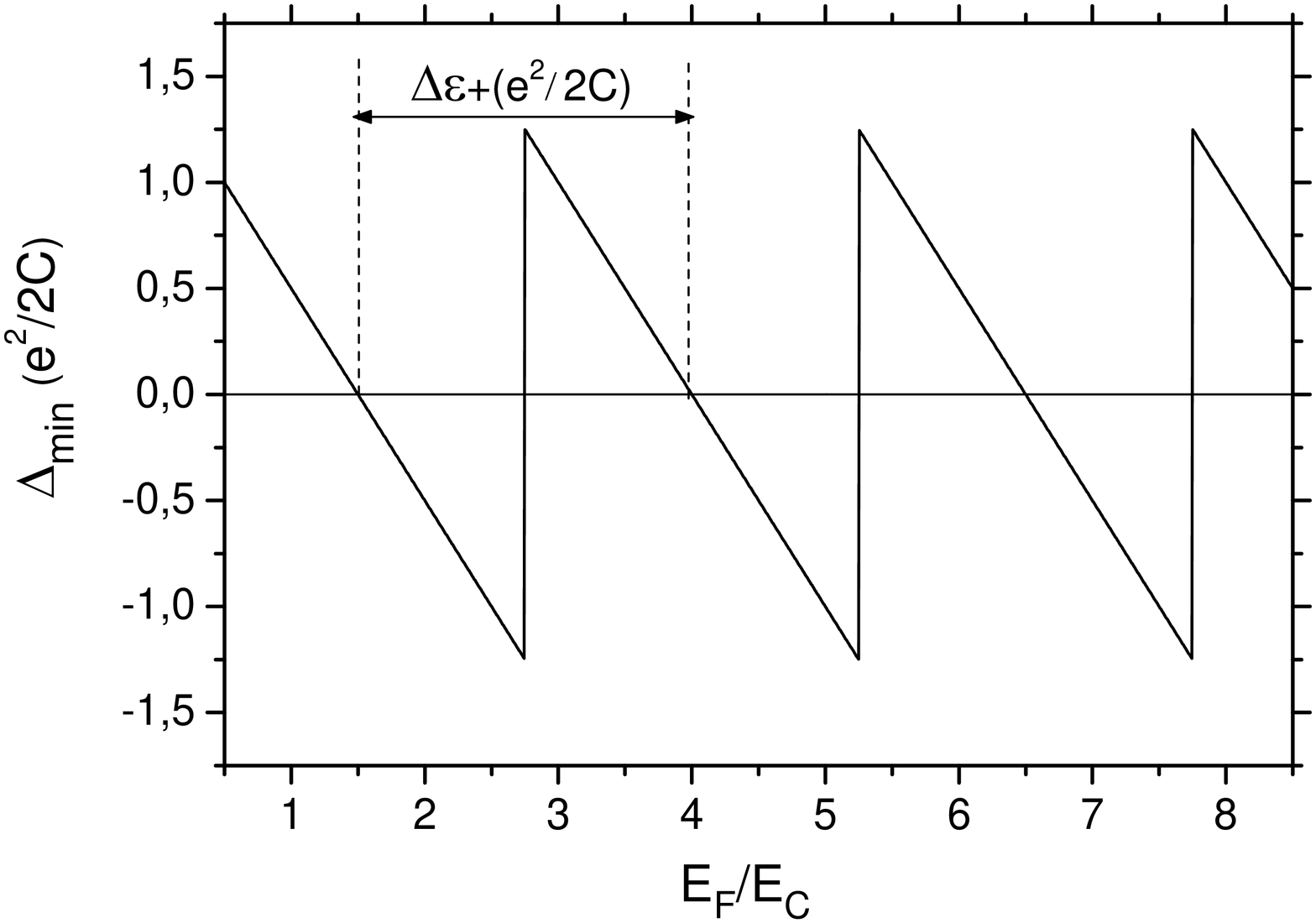}
\par\end{centering}

\caption{The Coulomb blockade oscillations of $\Delta_{min}$.
$\Delta\epsilon=0.5 E_{C}$.}
\end{figure}

The similarity between $-K$ and $S^{2}GT$ can be explained in
simple terms.  Namely, the dominant contribution to the summations
over $n$ and $N$ in Eq.~(\ref{sm}) occurs for $n=N=N_{min}$ where
$N_{min}$ is the integer which minimizes the absolute value of the
energy difference $\Delta(N)=\epsilon_{N}+U(N)-U(N-1)-E_{F}$. We
note that $\Delta(N_{min})$ oscillates in a sawtooth manner with
$E_{F}$ and passes through zero each time an electron enters the
dot. In what follows we keep Beenakker's notation \cite{Been1} and
we write $\Delta(N_{min})\equiv\Delta_{min}$. The oscillations of
$\Delta_{min}$ are shown in Fig.3.

In the low-$T$ limit the probabilities $P_{eq}(N_{min})$ and
$P_{eq}(\epsilon_{N_{min}}|N_{min})$ can be approximated by
\cite{Been1}
\begin{equation}
\label{p}
P_{eq}(N_{min})\approx f(\Delta_{min})
\end{equation}
and
\begin{equation}
P_{eq}(\epsilon_{N_{min}}|N_{min})\approx 1
\end{equation}
Inspection of Eq.~(\ref{sm}) shows that the quantities $s_{m}$ can
be written in the approximate form
\begin{equation}
\label{sma} s_{m}\approx
s_{m}^{N_{min}}=f(\Delta_{min})[1-f(\Delta_{min})](\Delta_{min})^{m},
\end{equation}
where the superscript $N_{min}$ denotes the contribution of the
term $n=N_{min}$ in the sum (\ref{sm}). Now, by substituting
Eq.~(\ref{sma}) into Eqs.(\ref{M}) and (\ref{K}) one readily gets
\begin{equation}
\label{Sap} S=\frac{M}{GT}=-\frac{1}{eT}\Delta_{min},
\end{equation}
and
\begin{equation}
\label{SGT} S^{2}GT=-K=\frac{\Gamma^{tot}}{k_{B}T^{2}}
f(\Delta_{min})[1-f(\Delta_{min})](\Delta_{min})^{2}.
\end{equation}
We recall that $f(\Delta_{min})=1/[\exp(\beta\Delta_{min})+1]$.

By maximizing Eq.~(\ref{SGT}) with respect to $\Delta_{min}$ we
find that $-K$ and $S^{2}GT$ exhibit two symmetric peaks at
$\beta\Delta_{min}=\pm 2.4$ shown in Fig.2. The magnitude of the
peaks is
\begin{equation}
\label{SGTmax} (S^{2}GT)^{max}=0.44k_{B}\Gamma^{tot}.
\end{equation}
We note that the position of the two strong maxima are in
agreement with the predicted maximization of $S^{2}GT$ and $ZT$ in
materials with delta-like density of states \cite{Mahan}.
Moreover, according to Eq.~(\ref{Sap}), the absolute magnitude of
$S$ that corresponds to the optimization of $S^{2}GT$ is $2.4
k_{B}/e=207$~$\mu$V/K which is in absolute agreement with the
value predicted in \cite{Mahan}.

At this point we should remark that Eq.~(\ref{sma}) describes
accurately the magnitude and the shape of the physical quantities
$G$, $M$, $K$ and $S^{2}GT$. Moreover, the expression (\ref{Sap})
for $S$ is in very good agreement with the numerical result around
$\Delta_{min}=0$ within the energy range $|\Delta_{min}|\lesssim
\Delta E$. The contribution of the terms $n\neq N_{min}$ in
Eq.(\ref{sm}) is responsible for the fine structure on the
oscillations of $S$ that was predicted by Beenakker and Staring
\cite{Been2}. These terms do not affect $S^{2}GT$ and are not
discussed here for simplicity.

The strong competition between $-K$ and $S^{2}GT$ in
Eq.(\ref{kappa}) gives $\kappa_{e}=0$ if only the leading
contribution to the sum (\ref{sm}) for $n=N_{min}$ and $N=N_{min}$
is considered. The terms that correspond to $n=N_{min}\pm 1$
disturb the above cancellation and, as we show below, an
analytical expression for $\kappa_{e}$ can be obtained. Inspection
of Eqs.(\ref{sm}) and (\ref{p}) shows that these terms introduce
tiny corrections to the quantities $s_{m}$ of the form
\begin{eqnarray}
\label{scor} s_{m}^{N_{min}\pm 1}&=&P_{eq}(\epsilon_{N_{min}\pm
1}|N_{min})(\Delta_{min}\pm \Delta \epsilon)^{m}
\nonumber\\
&\times& f(\Delta_{min})[1-f(\Delta_{min}\pm \Delta \epsilon)]
\end{eqnarray}
where the superscripts $N_{min}\pm 1$ denote the contribution from
the terms $n=N_{min}\pm 1$. The probabilities
$P_{eq}(\epsilon_{N_{min}\pm 1}|N_{min})$ are given by
[\onlinecite{Been2}]
\begin{equation}
P_{eq}(\epsilon_{N_{min}+1 }|N_{min})\approx\exp(-\beta\Delta
\epsilon )
\end{equation}
and
\begin{equation}
P_{eq}(\epsilon_{N_{min}-1 }|N_{min})\approx 1.
\end{equation}

In the energy interval around $\Delta_{min}=0$, which is the
regime where $G$ exhibits the well known peaks \cite{Been1},
$s_{m}^{N_{min}\pm1}$ are smaller by a factor of
$\exp(-\beta\Delta \epsilon)$ compared to $s_{m}^{N_{min}}$ given
by Eq.~(\ref{sma}) (we note that $\beta\Delta\epsilon\gg 1$). This
explains why these corrections are unimportant for $G$
\cite{Been1}. To obtain $\kappa_{e}$ we use Eq.(\ref{kappa}) after
we substitute $K$, $S$ and $G$ from Eqs.(\ref{K}), (\ref{S}) and
(\ref{G}), respectively. Then we take
\begin{equation}
\label{k1} \kappa_{e}=\frac{\Gamma^{tot}}{k_{B}T^{2}}\left
[s_{2}-\frac{(s_{1})^{2}}{s_{0}}\right]
\end{equation}
We, now,  approximate $s_{m}$ by the sum
$s_{m}^{N_{min}}+s_{m}^{N_{min}-1}+ s_{m}^{N_{min}+1}$ and we
handle $s_{m}^{N_{min}\pm 1}$ as small quantities in order to
linearize expression (\ref{k1}) by keeping terms up to first order
in $\exp(-\beta\Delta \epsilon)$ and obtain
\begin{equation}
\label{kappas1} \kappa_{e}\approx
\kappa_{e}^{N_{min}}+\kappa_{e}^{N_{min}+1}+\kappa_{e}^{N_{min}-1}
\end{equation}
where,
\begin{equation}
\label{kappas2}
\kappa_{e}^{N_{min}}=\frac{\Gamma^{tot}}{k_{B}T^{2}}\left(s_{2}^{N_{min}}-
\frac{(s_{1}^{N_{min}})^2}{s_{0}^{N_{min}}}\right)=0
\end{equation}
and
\begin{eqnarray}
\label{kappas3} \kappa_{e}^{N_{min}\pm
1}&=&\frac{\Gamma^{tot}}{k_{B}T^{2}}[s_{2}^{N_{min}\pm
1}\nonumber\\&+&\frac{s_{1}^{N_{min}}}{(s_{0}^{N_{min}})^2}(s_{1}^{N_{min}}s_{0}^{N_{min}\pm
1}-2s_{0}^{N_{min}}s_{1}^{N_{min}\pm 1})].\nonumber\\
\end{eqnarray}
The substitution of the quantities $s_{m}^{N_{min}}$ and
$s_{m}^{N_{min}\pm 1}$ from Eqs.(\ref{sma}) and (\ref{scor}) gives
\begin{eqnarray}
\label{cor} \kappa_{e}^{N_{min}\pm 1}
&=&k_{B}\,\Gamma^{tot}P_{eq}(\epsilon_{N_{min}\pm 1}|N_{min})
(\beta\Delta \epsilon)^{2}\nonumber\\
&\times& f(\Delta_{min})[1-f(\Delta_{min}\pm \Delta \epsilon)]
\end{eqnarray}
It is apparent from the above analysis that $\kappa_{e}$ consists
of two contributions
\begin{equation}
\label{kappas} \kappa_{e}\approx
\kappa_{e}^{N_{min}+1}+\kappa_{e}^{N_{min}-1}.
\end{equation}
These contributions are shown as dashed-dotted and dotted lines in
Fig.4. We see that they are symmetric around $\Delta_{min}=0$ and
peak at $\Delta_{min}=-\Delta \epsilon/2$ and $\Delta \epsilon/2$.
We note that $\kappa_{e}^{N_{min}+1}$ and $\kappa_{e}^{N_{min}-1}$
are periodic as a function of $E_{F}$ with periodicity
$\Delta\epsilon + (e^{2}/C)$.
\begin{figure}
\begin{centering}
\includegraphics[angle=-90,width=7.3cm]{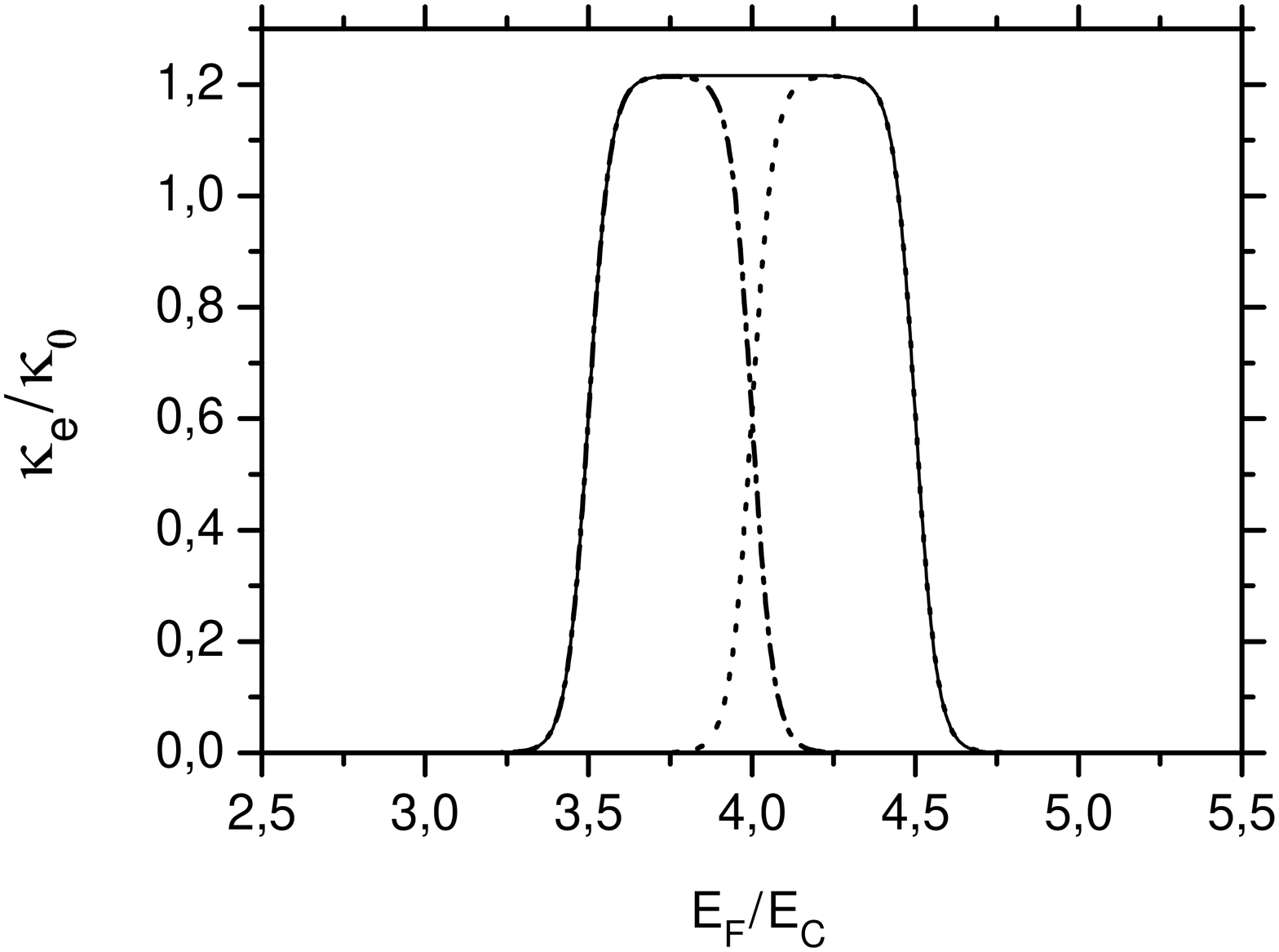}
\par\end{centering}

\caption{The two symmetric contributions to $\kappa_{e}$ (solid
line) for a dot with equidistant energy spectrum around $\Delta
_{min}=0$. The dashed-dotted and the dotted lines correspond to
$\kappa_{e}^{N_{min}-1}$ and $\kappa_{e}^{N_{min}+1}$,
respectively. $\Delta \epsilon=0.5 E_{C}$ and $\beta\Delta
\epsilon =15$. $\kappa_{0}=L_{0}TG^{max}(\beta\Delta
\epsilon)^{2}\exp(-\beta\Delta \epsilon)$ where $G^{max}=G_{0}/4$.
The shown structure is reproducible in energy intervals $\Delta
\epsilon+(e^{2}/C)$ except for the first and the last level where
$\kappa_{e}$ is $\kappa_{e}^{N_{min}+1}$ and
$\kappa_{e}^{N_{min}-1}$, respectively.}
\end{figure}

We should mention that there is a symmetry in the correction terms
$G^{N_{min}\pm 1}$ for the electrical conductance, arising from
the terms $n=N_{min}\pm 1$, and $\kappa_{e}^{N_{min}\pm 1}$.
Namely,
\begin{equation}
\kappa_{e}^{N_{min}\pm
1}=\left(\frac{k_{B}}{e}\right)^{2}T(\beta\Delta\epsilon)^{2}G^{N_{min}\pm
1}.
\end{equation}
Such a symmetry was absent in the papers \cite{Zianni2,Zianni1}.
This poses a question about the existence of a solid theoretical
background for the expressions for $\kappa_{e}$ that appeared in
these papers and about the validity of the relevant results
especially for the case of a dot with degenerate energy spectrum.

Substituting Eq.(\ref{cor}) in (\ref{kappas}), after some trivial
algebra we take \cite{Tsaousidou1,Tsaousidou2}
\begin{eqnarray}
\label{kappaf} \kappa_{e}&=&k_{B}\,\Gamma^{tot}
(\beta\Delta \epsilon)^{2}\nonumber\\
&\times&\frac{(1+e^{\beta\Delta \epsilon})e^{\beta\Delta_{min}}}
{e^{\beta\Delta
\epsilon}(e^{2\beta\Delta_{min}}+1)+e^{\beta\Delta_{min}}
(e^{2\beta\Delta \epsilon}+1)}.
\end{eqnarray}
Eq.~(\ref{kappaf}) is in excellent agreement with the numerical
result. The magnitude of the peak values of $\kappa_{e}$ around
$\Delta_{min}=0$ is given by the simple expression
\begin{equation}
\label{kappaa} \kappa_{e}\approx k_{B}\,\Gamma^{tot}(\beta\Delta
\epsilon)^{2}\exp(-\beta\Delta \epsilon).
\end{equation}
In deriving the above equation we have taken into account that
$\exp(\beta\Delta \epsilon)\gg 1$ in the quantum limit.
Eq.~(\ref{kappaa}) denotes a strong sensitivity of $\kappa_{e}$ to
the details of the energy spectrum.

Inspection of Eqs.(\ref{G}) and (\ref{kappaa}) shows that the
Wiedemann-Franz law is not valid in the quantum limit. We find
that the peak values of the Coulomb blockade oscillations of
$\kappa_{e}$ are smaller than that the Wiedemann-Franz law
predicts by a factor of the order $(\beta\Delta
\epsilon)^{2}\exp(-\beta\Delta\epsilon)$. Namely, the ratio of the
peak values of $\kappa_{e}$ and $G$ for $\Delta_{min}=0$ is given
by
\begin{equation}
\frac{\kappa_{e}^{max}}{G^{max}}=\frac{12}{\pi^{2}}L_{0}T(\beta\Delta
\epsilon)^{2}\exp(-\beta\Delta\epsilon).
\end{equation}
As we show below the violation of the Wiedemann-Franz law in the
system under consideration leads to an exponential increase of
$ZT$ with $\beta\Delta\epsilon$. Also, recently it has been
reported that deviations from the Wiedemann-Franz law lead to
enhancement of ZT in nanocontacts made of two-capped single wall
nanotubes~\cite{Esfarjani}, quantum dots attached to ferromagnetic
leads~\cite{Swirkowicz}, and strongly correlated quantum
dots\cite{Costi}.

The maximization of $S^{2}GT$, discussed earlier, has a direct
impact on the dimensionless figure of merit $ZT$ when the phonon
contribution to the thermal conductance is ignored
\begin{equation}
ZT=\frac{S^{2}GT}{\kappa_{e}}.
\end{equation}
According to the above expression and Eqs.(\ref{SGTmax}) and
(\ref{kappaa}) $ZT$ shows two strong maxima at
$\beta\Delta_{min}=\pm 2.4$ of magnitude
\begin{equation}
(ZT)^{max}=
\frac{(S^{2}GT)^{max}}{\kappa_{e}}=0.44\exp(\beta\Delta
\epsilon)/(\beta\Delta \epsilon)^{2}.
\end{equation}
The structure of $ZT$ is shown in Fig.5c. The exponential
dependence of these maxima can be readily explained by the fact
that, in the energy interval where $S^{2}GT$ exhibits the peak
structure shown in Fig.5a, $\kappa_{e}$ remains constant and its
magnitude is given by Eq.~(\ref{kappaa}) (see Fig.5b).

Eq.~(30) shows that the thermoelectric efficiency can be
dramatically enhanced by controlling the size of the dot and
consequently $\Delta\epsilon$. For small dots (or molecules) the
requirement that the system is in the quantum limit is fulfilled
even at room temperature. Practically speaking, the quantum limit
is approached when $\beta\Delta\epsilon\geq 5$. High values of
$ZT$ in the range of 2.6 to 97 at $T=300$~K ($k_{B}T=0.026$~eV)
can be obtained when $\Delta\epsilon$ varies between 5 to 10
$k_{B}T$ (namely, between 0.13 to 0.26~eV). We note that the
energy separation between the first two energy levels in a
spherical dot for a material with electron effective mass of the
order of $0.1~m_{e}$ ($m_{e}$ is the electron mass) is greater
than 0.15~eV when the diameter is less than 5~nm. Finally, we add
that in our model we have assumed that the charging energy $E_{C}$
is larger than $\Delta\epsilon$. Large values of $E_{C}$ of the
order of the tenth of eV have been reported recently in
single-electron transistors with organic molecules acting as
Coulomb islands\cite{Kubota1,Kubota2}.

\begin{figure}
\begin{centering}
\includegraphics[angle=0,height=9.5cm]{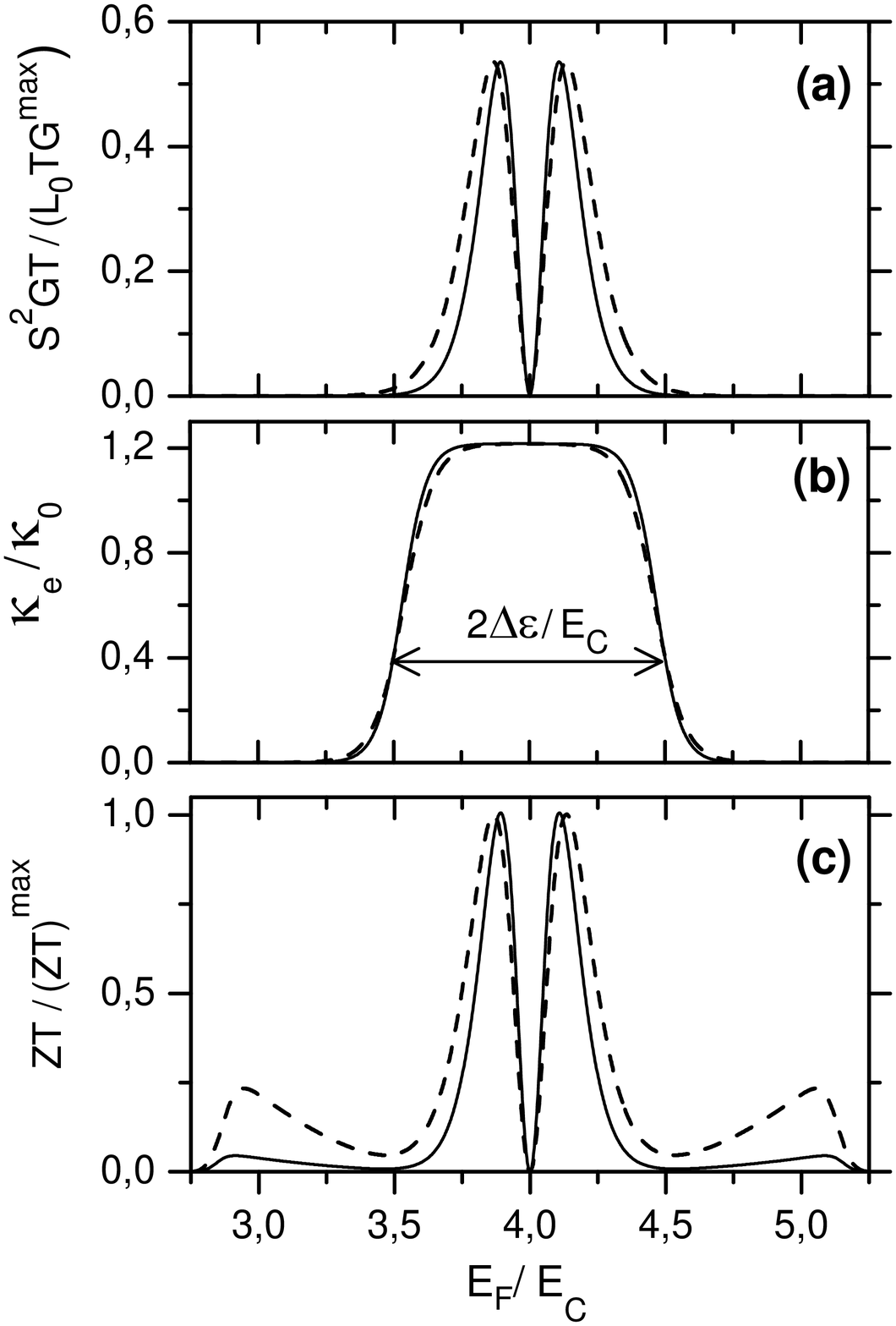}
\par\end{centering}

\caption{Calculated values of $S^{2}GT$, $\kappa_{e}$ and $ZT$ as
a function of $E_{F}$ for a three-level dot. The level spacing is
$\Delta \epsilon=0.5 E_{C}$. The solid and dashed lines
correspond, respectively, to $\beta\Delta \epsilon$=11 and 9. The
results are plotted around $E_{F}=4E_{C}$ that corresponds to
$\Delta_{min}=0$ when an electron enters the dot at the level
$n=2$. In (b) $\kappa_{0}=L_{0}TG^{max}(\beta\Delta
\epsilon)^{2}\exp(-\beta\Delta \epsilon)$ where $G^{max}=G_{0}/4$
and in (c) $(ZT)^{max}=0.44\exp(\beta\Delta \epsilon)/(\beta\Delta
\epsilon)^{2}$.}
\end{figure}

\begin{figure}
\begin{centering}
\includegraphics[angle=0,height=9.5cm]{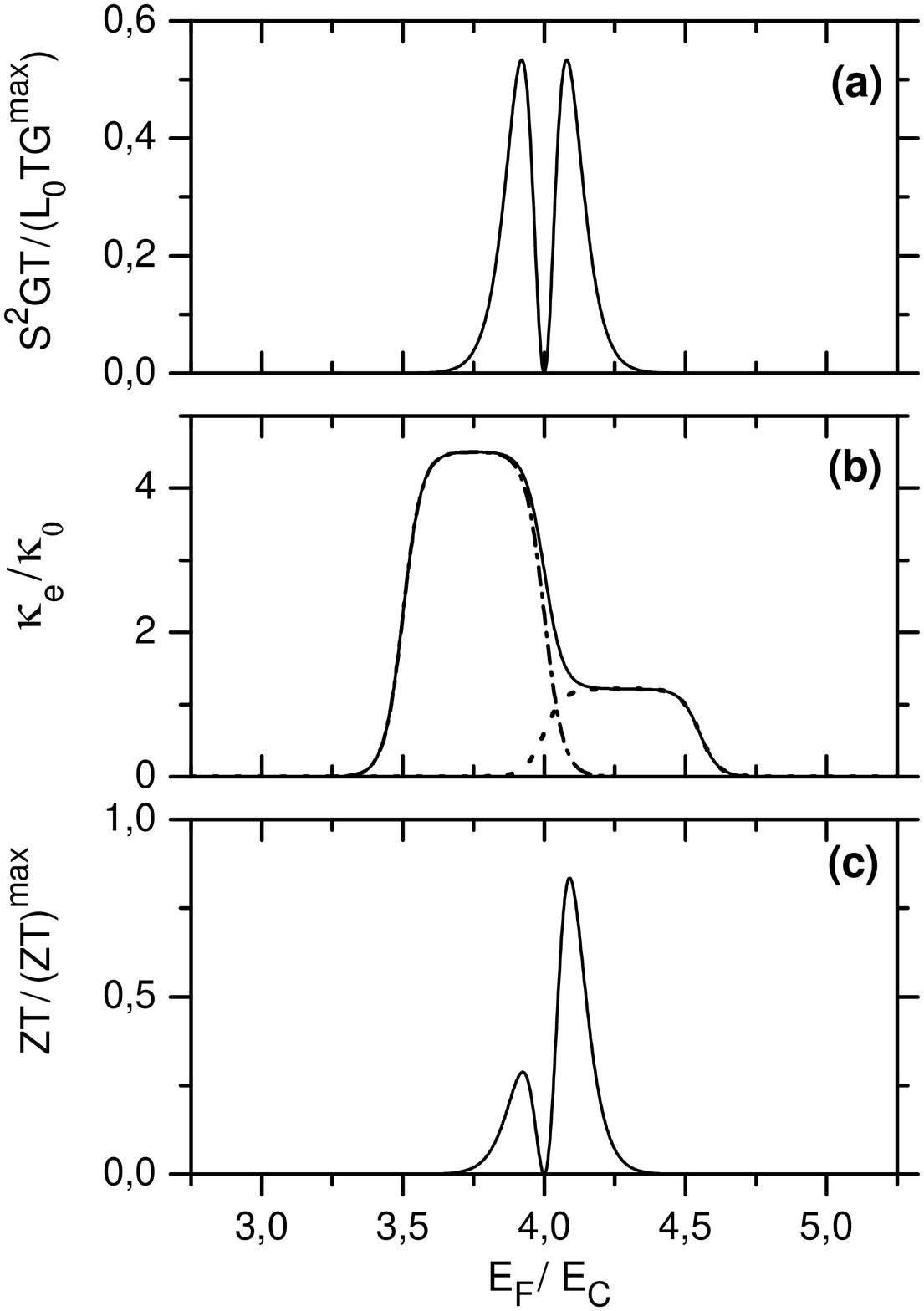}
\par\end{centering}

\caption{Calculated values of $S^{2}GT$, $\kappa_{e}$ and $ZT$ as
a function of $E_{F}$ for a dot with non-equidistant energy
spectrum. The results are plotted around $E_{F}=4E_{C}$ that
corresponds to $\Delta_{min}=0$ when an electron enters the dot at
the level $n=2$. The energy separation between the two adjacent
levels is $\Delta \epsilon_{pr}=0.5 E_{C}$ and $\Delta
\epsilon_{ne} =0.55 E_{C}$. $\beta\Delta \epsilon_{pr}=15$ and
$\beta\Delta \epsilon_{ne}=16.5$. The dashed-dotted and the dotted
lines in (b) show the contributions $\kappa_{e}^{N_{min}-1}$ and
$\kappa_{e}^{N_{min}+1}$, respectively.
$\kappa_{0}=L_{0}TG^{max}(\beta\Delta
\epsilon_{ne})^{2}\exp(-\beta\Delta \epsilon_{ne})$ and
$(ZT)^{max}=0.44\exp(\beta\Delta \epsilon_{ne})/(\beta\Delta
\epsilon_{ne})^{2}$.}
\end{figure}

In a recent paper \cite{Zianni1} the possibility of tuning $ZT$ of
a dot by varying $E_{F}$ was recognized. In [\onlinecite{Zianni1}]
no detailed calculations were performed and the possibility of
tuning $ZT$ was incorrectly related solely to the dependence of
$\kappa_{e}$ on the energy spectrum and the variation of
$\kappa_{e}$ with $E_{F}$. However, the variation of $S^{2}GT$
with $E_{F}$ is the most remarkable behavior predicted in these
systems. Although, the phonon contribution to the thermal
conductance could washout the variation of the total thermal
conductance as a function of $E_{F}$, $ZT$ will still vary
periodically with $E_{F}$ following the structure of $S^{2}GT$
shown in Figs.2 and 5a. Moreover, the maximization of $S^{2}GT$
could be utilized for the prediction of an optimal value for
$\kappa_{ph}$ that maximizes $ZT$ \cite{Murphy}.

So far our analysis has referred to a dot with equidistant energy
levels. The generalization to a non-equidistant energy spectrum is
straightforward. However, we should remark that this
generalization will not affect our findings for $S^{2}GT$ (apart
from the absence of a regular periodicity in the oscillations of
$S^{2}GT$) but it will only affect the shape of $ZT$ in the case
where the phonon contribution to thermal conductance is small
compared to $\kappa_{e}$. This is due to the profound effect of
the energy spectrum on the shape of $\kappa_{e}$. In Fig.6 we
present calculations for $S^{2}GT$, $\kappa_{e}$ and $ZT$ for a
three-level dot when one electron enters the dot at the level
$n=2$. The energy separation between the next and the previous
energy level is $\Delta \epsilon_{ne}=0.55 E_{C}$ and $\Delta
\epsilon_{pr}=0.5 E_{C}$. To obtain $\kappa_{e}$ in the case of
dot with non-equidistant energy spectrum one needs to replace
$\Delta \epsilon$ by $\Delta \epsilon_{ne}$ and $\Delta
\epsilon_{pr}$ in the expressions (\ref{cor}) for
$\kappa_{e}^{N_{min}+1}$ and $\kappa_{e}^{N_{min}-1}$,
respectively. In this case the two contributions to $\kappa_{e}$
become asymmetric as shown by the dashed-dotted and dotted lines
in Fig.6b. This results to the two asymmetric peaks in $ZT$ shown
in Fig.6c.

We should remark at this point that the phonon contribution to
thermal conductance could significantly suppress $ZT$. However,
the phonon confinement in free-standing QDs or QDs embedded in a
matrix material, where there is a significant mismatch in the
elastic properties of the QD and the surrounding material, could
dramatically decrease the phonon contribution $\kappa_{ph}$. The
3D confinement of phonons in nano-objects results in similar
restrictions in the phase space of the phonon wave vector as in
the case of electrons. The discrete phonon frequency spectrum
influences the phonon heat transfer through a nanostructure. In
what follows we will attempt to reveal the similarities between
the fermionic and the bosonic contributions to the thermal
conductance. As we will show both contributions show an activated
behavior.

We start from the Landauer expression used by Rego and Kirczenow
\cite{Rego} for the heat transferred by 1D ballistic phonons
through a quantum wire attached to two thermal reservoirs in the
presence of a small temperature difference $\Delta T$ between
them. This is modified here to account for the discrete phonon
spectrum of the QD. We assume that the temperature in the right
reservoir is raised by $\Delta T$ compared to the temperature in
the left reservoir. Then the phonon contribution to the heat flow
from the right to the left reservoir is written in the form
\cite{Ciraci}
\begin{equation}
Q_{ph}=\sum_{i}\int d\omega \hbar\omega
T_{i}(\omega)[n_{R}(\omega)-n_{L}(\omega)]
\end{equation}
where $T_{i}(\omega)$ denotes the transmission coefficient for a
phonon of frequency $\omega$ through the $i$th-mode into the
nanostructure and
$n_{R}(\omega)=\{\exp[\frac{\hbar\omega}{k_{B}(T+\Delta
T)}]-1\}^{-1}$ and
$n_{L}(\omega)=[\exp(\frac{\hbar\omega}{k_{B}T})-1]^{-1}$ are the
phonon distribution functions in the right and the left
reservoirs, respectively. The summation is over all the
vibrational modes. For the case of resonance energy transfer we
write the transmission coefficient $T_{i}(\omega)$ in the
form~\cite{Segal}
\begin{equation}
T_{i}(\omega)
=\frac{\Gamma^{L}_{ph}\Gamma^{R}_{ph}}{\Gamma^{L}_{ph}+\Gamma^{R}_{ph}}
\delta(\omega-\omega_{i})
\end{equation}
where $\Gamma_{ph}$ denotes the phonon tunnel rate. Then we take
\begin{equation}
\label{heatph}
Q_{ph}=\sum_{i}\hbar\omega_{i}\Gamma^{tot}_{ph}[n_{R}(\omega_{i})-n_{L}(\omega_{i})],
\end{equation}
where $\Gamma_{ph}^{tot}=\Gamma^{L}_{ph}\Gamma^{R}_{ph}/
(\Gamma^{L}_{ph}+\Gamma^{R}_{ph})$. The expression (\ref{heatph})
is in agreement with the expression derived recently by Segal and
Nitzan~\cite{Segal} for the heat current through a harmonic
molecule weakly coupled to two thermal reservoirs by using a
master equation approach.

In the linear regime, where $\Delta T$ is small, it can be easily
shown that
\begin{equation}
n_{R}(\omega_{i})-n_{L}(\omega_{i})=-\hbar\omega_{i}\frac{\partial
n_{L}(\omega_{i})}{\partial (\hbar\omega_{i})}\frac{\Delta T}{T},
\end{equation}
where,
\begin{equation}
\frac{\partial n_{L}(\omega_{i})}{\partial
(\hbar\omega_{i})}=-\beta\frac{\exp(\beta\hbar\omega_{i})}
{[\exp(\beta\hbar\omega_{i})-1]^{2}}.
\end{equation}
At temperatures where $\beta\hbar\omega_{i}>>1$ we can write
\begin{equation}
\frac{\partial n_{L}(\omega_{i})}{\partial
(\hbar\omega_{i})}\approx -\beta\exp(-\beta\hbar\omega_{i})
\end{equation}
Then, the phonon contribution to the thermal conductance
$\kappa_{ph}=Q_{ph}/\Delta T$ is written in the form
\begin{equation}
\kappa_{ph}=k_{B}\sum_{i}\Gamma_{ph}^{tot}(\beta\hbar\omega_{i})^{2}
\exp(-\beta\hbar\omega_{i}).
\end{equation}

At low $T$ only the first low energy phonon mode $\hbar\omega_{0}$
contributes significantly to $\kappa_{ph}$. Then we can write
\begin{equation}
\label{kappaph}
\kappa_{ph}=k_{B}\Gamma^{tot}_{ph}(\beta\hbar\omega_{0})^{2}\exp(-\beta\hbar\omega_{0}).
\end{equation}
The similarity of the above expression with Eq.(\ref{kappaa}) for
$\kappa_{e}$ is quite remarkable. Interestingly, as in the case of
the 1D thermal conductance~\cite{Rego}, the phononic and the
electronic contributions to the thermal conductance in structures
with a 3D electron and phonon confinement appear to show a similar
$T$-behavior.

According to our analysis the maximum value for $ZT$ is
\begin{equation}
\label{max} (ZT)^{max}=\frac{0.44
k_{B}\Gamma^{tot}}{\kappa_{e}+\kappa_{ph}}
\end{equation}
where $\kappa_{e}$ and $\kappa_{ph}$ are given by
Eqs.(\ref{kappaa}) and (\ref{kappaph}). Due to the activated terms
in these equations $ZT$ can take values much larger than 1. Our
finding suggests that the control of the electron and phonon
confinement effects in nanostructures can improve significantly
their thermoelectric properties. Namely, for $\hbar\omega_{0}$ and
$\Delta\epsilon$ of the order of $150$ to $200$~meV and
$\Gamma^{tot}\approx\Gamma^{tot}_{ph}$ we find that $ZT$ can reach
the values 2.2 to 8.4 at $T=300$~K. We also note that $(ZT)^{max}$
is an increasing function of $\Gamma^{tot}/\Gamma^{tot}_{ph}$.

\section{Conclusions}
In summary, we have presented a comprehensive study of the
thermoelectric coefficients of a multi-level QD weakly coupled to
two electron reservoirs in the Coulomb blockade regime for
sequential tunneling. Our analysis focuses on the quantum
limit\cite{Been1} where the level spacing between successive
electronic levels is much larger than the thermal energy
($\beta\Delta\epsilon>>1$). Analytical expressions for the power
factor and the figure of merit are derived. We found an
exponential increase of ZT with $\beta\Delta\epsilon$ when phonons
are ignored. This is due to the activated behavior of
$\kappa_{e}$. We also show that $\kappa_{ph}$ for a dot with
discrete phonon spectrum shows a similar activated behavior.
Interestingly both fermionic and bosonic contributions to the
thermal conductance show a very similar $T$-dependence. Although,
it was previously believed that electron-transmitting and
phonon-blocking structures have good thermoelectric properties
\cite{Venka}, our results now open a route for designing
electron-blocking and phonon-blocking nanostructures with improved
thermoelectric performance.


\begin{thebibliography}{99}
\bibitem{Dresselhaus} M. S. Dresselhaus, G. Chen, M. Y. Tang, R. G. Yang,
H. Lee, D. Z. Wang, Z. F. Ren, J.-P. Fleurial, and P. Gogna, Adv.
Mater. {\bf 19}, 1043 (2007).
\bibitem{Majumdar} A. Majumdar, Science {\bf 303}, 777 (2004).
\bibitem{Venka} R. Venkatasubramanian, E. Siivola, T. Colpitts, and
B. O'Quinn, Nature {\bf 413}, 597 (2001).
\bibitem{Harman} T. C. Harman, P. J. Taylor, M. P. Walsh, and B. E.
LaForge, Science {\bf 297}, 2229 (2002).
\bibitem{Hsu} K. F. Hsu, S. Loo, F. Guo, W. Chen, J. S. Dyck, C.
Uher, T. Hogan, E. K. Polychroniadis, and M. G. Kanatzidis,
Science {\bf 303}, 818 (2004).
\bibitem{Poudel} B. Poudel, Q. Hao, Y. Ma, Y. Lan, A. Minnich, B.
Yu, X. Yan, D. Wang, A. Muto, D. Vashaee, X. Chen, J. Liu, M. S.
Dresselhaus, G. Chen, and Z. Ren, Science {\bf 320}, 634 (2008).
\bibitem{Hochbaum} A. I. Hochbaum, R. Chen, R. D. Delgado, W. Liang,
E. C. Garnett, M. Najarian, A. Majumdar, and P. Yang, Nature {\bf
451}, 163 (2008).
\bibitem{Boukai} A. I. Boukai, Y. Bunimovich, J. Tahir-Kheli, J. -K.
Yu, W. A. Goddard III, and J. R. Heath, Nature {\bf 451}, 168
(2008).
\bibitem{Fletcher} R. Fletcher, E. Zaremba, and U. Zeitler in {\em Electron-Phonon
Interactions in Low Dimensional Structures}, edited by L. Challis
(Oxford University Press, Oxford, 2003), p. 149.
\bibitem{Tsaousidou} M. Tsaousidou in {\em The Oxford Handbook of
Nanoscience and Technology}, edited by A.V. Narlikar and Y.Y. Fu,
Vol.II, Ch.13, (Oxford University Press, Oxford, 2010), p.477.
\bibitem{Hicks1} L. D. Hicks and M. S. Dresselhaus, Phys. Rev. B {\bf
47}, 12727 (1993); L. D. Hicks and M. S. Dresselhaus, Phys. Rev. B
{\bf 47}, 16631 (1993).
\bibitem{Mahan} G. D. Mahan and J. O. Sofo, Proc. Natl. Acad. Sci.
USA {\bf 93}, 7436 (1996).
\bibitem{Humphrey} T. E. Humphrey and H. Linke, Phys. Rev. Lett.
{\bf 94}, 096601 (2005).
\bibitem{Tsaousidou1}  M. Tsaousidou and G. P. Triberis,
arXiv:cond-mat/0605286 (May 2006).
\bibitem{Tsaousidou2} M. Tsaousidou and G. P. Triberis, in {\em
Physics of Semiconductors: 28th International Conference on the
Physics of Semiconductors-ICPS 2006}, AIP Conf. Proc. {\bf 893}
(AIP, New York 2007) p. 801-802.
\bibitem{Murphy} P. Murphy, S. Mukerjee, and J. Moore,
Phys. Rev. B {\bf 78}, 161406(R) (2008).
\bibitem{Lambert} C. M. Finch, V. M. Garc\'{i}a-Su\'{a}rez, and C. J.
Lambert, Phys. Rev. B {\bf 79}, 033405 (2009).
\bibitem{Swirkowicz} R. \'{S}wirkowicz, M. Wierzbicki, and J. Barna\'{s},
Phys. Rev. B {\bf 80}, 195409 (2009).
\bibitem{Zianni2} X. Zianni, Phys. Rev. B {\bf 75}, 045344 (2007).
\bibitem{Kubala} B. Kubala, J. K\"{o}nig, and J. Pekola, Phys. Rev. Lett. {\bf 100},
066801 (2008).
\bibitem{Been1} C. W. J. Beenakker, Phys. Rev. B {\bf 44}, 1646
(1991).
\bibitem{Been2} C. W. J. Beenakker and A. A. M. Staring, Phys. Rev. B {\bf 46},
9667 (1992).
\bibitem{Zianni1} X. Zianni, Phys. Rev. B {\bf 78}, 165327 (2008).
\bibitem{Esfarjani} K. Esfarjani, M. Zebarjadi, and Y. Kawazoe,
Phys. Rev. B {\bf 73}, 085406 (2006).
\bibitem{Costi} T. A. Costi and V. Zlati\'{c}, Phys. Rev. B {\bf 81}, 235127
(2010).
\bibitem{Kubota1} Y. Wakayama, T. Kubota, H. Suzuki, T. Kamikado,
and S. Mashiko, Journal of Appl. Phys. {\bf 94}, 4711 (2003).
\bibitem{Kubota2} Y. Wakayama, T. Kubota, H. Suzuki, T. Kamikado,
and S. Mashiko, Nanotechnology {\bf 15}, 1446 (2004).
\bibitem{Rego} L. G. C. Rego and G. Kirczenow, Phys. Rev. Lett. {\bf
81}, 232 (1998).
\bibitem{Ciraci} A. Ozpineci and S. Ciraci, Phys. Rev. B {\bf 63},
125415 (2001).
\bibitem{Segal} D. Segal and A. Nitzan, Phys. Rev. Lett. {\bf 94},
034301 (2005).




\end{thebibliography}
\end{document}